\documentclass[journal=langd5,manuscript=article]{achemso}

\usepackage[version=3]{mhchem} 
\usepackage{stix}
\usepackage{xcolor}
\usepackage{siunitx}
\usepackage{graphicx}

\def \rb {\mathbf{r}}
\author{Christopher Kessler}
\author{Johannes Eller}
\author{Joachim Gross}
\author{Niels Hansen}
\email{hansen@itt.uni-stuttgart.de}
\affiliation[University of Stuttgart]
{Institute of Thermodynamics and Thermal Process Engineering, University of Stuttgart, Pfaffenwaldring 9, 70569 Stuttgart, Germany}

\title[An \textsf{achemso} demo]
{Adsorption of Light Gases in Covalent Organic Frameworks: Comparison of Classical Density Functional Theory and Grand Canonical Monte Carlo Simulations}

\abbreviations{GCMC, DFT, COFs, COF}

\begin{document}






\begin{abstract}
    A classical density functional theory (cDFT) based on the PC-SAFT equation of state is proposed for the calculation of adsorption equilibrium of pure substances and their mixtures in covalent organic frameworks (COFs).
    Adsorption isotherms of methane, ethane, \textit{n}-butane and nitrogen in the COFs TpPa-1 and 2,3-DhaTph are calculated and compared to results from grand canonical Monte Carlo (GCMC) simulations. Mixture adsorption is investigated for the methane/ethane and methane/\textit{n}-butane binary systems. Excellent agreement between PC-SAFT DFT and GCMC is obtained for all adsorption isotherms up to pressures of 50~bar. The cDFT formalism accurately predicts the selective accumulation of longer hydrocarbons for binary mixtures in the considered COFs. This application shows substantial predictive power of PC-SAFT DFT solved in three-dimensional geometries and the results suggest the method can in the future also be applied for efficient optimization of force field parameters or of structural properties of the porous material based on analytical theory as opposed to a stochastic simulation.
\end{abstract}

\section{Introduction}
Covalent organic frameworks (COFs) are ordered nanoporous materials formed by covalent bonds between organic building blocks composed of light elements, such as carbon, nitrogen, oxygen, and hydrogen\cite{Cote_2005}.
The materials are characterized by their large surface area, high porosity and low molecular weights.
Therefore, a broad variety of applications have been envisioned, including gas storage and separation, catalysis, sensing, drug delivery, and optoelectronic materials development\cite{Huang_Angew_2015,Huang_2016,Song_Adv_Sci_2019,Li_IECR_2019,C9MH00856J,Chen_Angew_2020,C9CS00827F,Gottschling_JACS_2020,Nagai_book_2020}.
Their bottom-up synthesis based on small building blocks allows the design of porous materials possessing a large variety of pore sizes and topologies. 
Similar to other porous materials such as zeolites or metal organic frameworks (MOFs), the number of hypothetical structures exceeds the ones synthesized so far by three orders of magnitude\cite{Lan_2018}. Databases of curated structures\cite{TONG2017456,TONG2018,Ongari_ACSCS_2019} and computational workflows that automatize molecular simulation and analysis are being developed to screen materials for different purposes\cite{Yan_ACSSCE_2019,Deeg_ACSAMI_2020,Ongari_ACSCS_2020}.

In two-dimensional (2D) COFs the organic building blocks are linked into 2D atomic layers that further 
stack via $\pi$-$\pi$ interactions to crystalline layered structures. The manner in which adjacent sheets stack in this assembly process forming the crystalline material largely 
influence their material properties including pore accessibility and, in turn, adsorption capacity\cite{Putz_2020,Mahringer_2020}.
It is therefore unclear how representative idealized structural models can be compared to real COF materials. 
This calls for an efficient computational approach that is able to quantify the impact of structural variations on the adsorption behavior. An established technique
for this purpose are molecular simulations, in particular Monte Carlo simulations in the grand canonical ensemble\cite{Norman_1969} (GCMC).
Molecular simulation studies targeting adsorption and/or diffusion in COFs have considered relatively small adsorbate molecules such as helium, argon, hydrogen, methane, nitrogen or carbon dioxide,\cite{Garberoglio_2007,Babarao_2008,Liu_IECR_2010,Mendoza_Cortes_JPCA_2010,Zeng_Mol-Simul_2018,Keskin_JPCC_2012,C7CE01647F,Sharma_IECR_2018} respectively, for which force fields can be expected to reproduce the fluid properties with reasonable accuracy. However, for CO$_2$-adsorption on all-silica zeolites it was shown that computed Henry coefficients may differ by more than two orders of magnitude across different CO$_2$ force fields, in particular for zeolites with more confined pore features, while different force fields yield consistent predictions of Henry coefficients, when structures are less confined.\cite{Lim_2018}
In the case of COFs, containing significantly larger pore sizes compared to zeolites, the impact of the stacking motifs of adjacent layers in the structural model used in the simulations is expected to influence the simulation outcome at least to the same extent as residual discrepancies in the force fields used.\cite{Garberoglio_2007,Keskin_JPCC_2012,C7CE01647F,Sharma_IECR_2018}

To increase versatility of computational methods, a more efficient alternative to molecular simulation could be classical density functional theory (cDFT)\cite{Evans_1979,Wu_2017} which is also rooted in the framework of statistical mechanics but relies on an inhomogeneous density profile compared to explicit atomistic molecular simulation. 
One of the most common application of cDFT is adsorption in homogeneous slit pores with two opposing planar walls. The solid is thereby modeled by an external field which commonly takes the form of a Lennard-Jones 9-3 potential or a Steele potential.
cDFT accurately predicts the adsorption behaviour when compared to GCMC simulations including layering transitions \cite{peterson1990layering} and capillary condensation \cite{sauer2019}. 
The adsorption behaviour of real unordered porous materials, however, is often not well represented by the homogeneous slit pore model with one given pore size.
This is because of the ambiguous pore structures with often unknown porosity, chemical composition and pore size distributions.
Therefore, cDFT models include heterogeneities \cite{shen2014modeling}, both in pore size distribution and surface roughness/chemical heterogeneity, to compare accurately to adsorption experiments.
Ordered porous media, in turn, are characterized by their regular pore structure and, thus, provide a consistency test between cDFT and molecular simulations beyond one-dimensional homogeneous slit pores.

A key ingredient of cDFT is the Helmholtz energy functional used to describe the fluid-fluid interactions. 
Whereas the hard-sphere repulsion is often represented by a functional based on Fundamental Measure Theory\cite{rosenfeld1989free, roth2002fundamental, yu2002structures}, dispersive attractions are either treated by a simple mean-field theory which ignores density correlations of the fluid or non-local weighted density approximations in combination with an underlying equation of state. 
Comparative computational studies of the adsorption in ordered porous frameworks between cDFT and GCMC simulations were performed by different groups; each utilizing different Helmholtz energy functionals. Guo and coworkers \cite{guo2018fast} compared adsorption isotherms of noble gases in metal organic frameworks using a mean-field approach. Fu and Wu \cite{fu2015classical} assessed the performance of different dispersive Helmholtz energy functionals from mean-field theory to weighted density approximations with an empirical equation of state. They compared calculated adsorption isotherms of methane in metal organic frameworks to molecular simulations.

This study uses a functional based on the Perturbed-Chain Statistical Associating Fluid Theory (PC-SAFT) equation state\cite{GrossSadowski2000, GrossSadowski2001}, which also utilizes a weighted density approximation\cite{sauer2017classical}. This functional was already successfully applied to adsorption in one-dimensional slit pores\cite{sauer2019} and the calculation of surface tension and Tolman lengths.\cite{rehner2018surface}

In this study we assess the PC-SAFT DFT model for predicting adsorption in ordered three-dimensional COF frameworks.
We consider the adsorption of light gases in two typical COFs and we compare results from GCMC and cDFT. The results are discussed in light of methodological differences of the two approaches.

\section{Computational Details}

\subsection{Classical Density Functional Theory}
In this section, we summarize the fundamental equations of classical density functional theory and the application to adsorption in COFs.
Density functional theory is formulated in the grand canonical ensemble at constant chemical potentials $\boldsymbol{\mu}=\left\lbrace \mu_i,i=1,\dots,\nu\right\rbrace$ of all species, volume $V$, and temperature $T$. The grand canonical potential was shown to be a unique functional of the inhomogeneous density profile $\boldsymbol{\rho}\left(\rb\right)=\left\lbrace \rho_i(\rb),i=1,\dots,\nu\right\rbrace$ and can be expressed as
\begin{equation}
    \Omega\left[\boldsymbol{\rho}\left(\rb\right)\right]=F\left[\boldsymbol{\rho}\left(\rb\right)\right]-\sum_{i=1}^\nu \int \rho_i \left(\mu_i-V_i^\mathrm{ext}\left(\rb\right) \right)\mathrm{d} \rb
\end{equation}
where $F\left[\boldsymbol{\rho}\left(\rb\right)\right]$ is the intrinsic Helmholtz energy functional capturing the fluid-fluid interactions and $V_i^\mathrm{ext}\left(\rb\right)$ is the external potential due to solid-fluid interactions acting on species $i$. For adsorption in microporous materials it is instructive to think of the system as being connected to a large bulk reservoir with the same temperature and chemical potentials $\boldsymbol{\mu}$, so that a pressure of a communicating bulk fluid $p(\boldsymbol{\mu},T)$ can be calculated.
The equilibrium density distribution $\boldsymbol{\rho}^0\left(\rb\right)$ minimizes the grand canonical functional
\begin{equation}
    \Omega\left[\boldsymbol{\rho}\left(\rb\right)\neq\boldsymbol{\rho}^0\left(\rb\right)\right]>\Omega\left[\boldsymbol{\rho}^0\left(\rb\right)\right]=\Omega\left(\boldsymbol{\mu},V,T\right)
\end{equation}
and its value is then equal to the grand canonical potential $\Omega\left(\boldsymbol{\mu},V,T\right)$, so that
\begin{equation}
    \left.\frac{\delta \Omega\left[\boldsymbol{\rho}\right]}{\delta \rho_i}\right|_{\rho_i(\rb)=\rho_i^0\left(\rb\right)}=0  \qquad\qquad \forall i
\end{equation}
The equilibrium density profile is obtained by solving the Euler-Lagrange equation
\begin{equation}
    \frac{\delta \Omega\left[\boldsymbol{\rho}(\rb)\right]}{\delta \rho_i}=\frac{\delta F\left[\boldsymbol{\rho}(\rb)\right]}{\delta \rho_i(\rb)}-\mu_i+V_i^\mathrm{ext}\left(\rb\right)=0
    \label{eq:euler_lagrange}
\end{equation}
using a damped Picard iteration in combination with an Anderson mixing scheme to accelerate the convergence rate\cite{mairhofer2017numerical}. For a compact notation, we henceforth omit the subscript $0$ in the equilibrium density profile; we use $\boldsymbol{\rho}\left(\rb\right)$ for the vector of density profiles of all components in the system.
The intrinsic Helmholtz energy functional $F\left[\boldsymbol{\rho}(\rb)\right]$ describes the fluid-fluid interactions and is based on the PC-SAFT equation of state. The coarse-grained molecular model of the PC-SAFT equation of state represents molecules as chains of tangentially bound spherical segments. In this work, we only consider non-polar, non-associating molecules, leading to the following Helmholtz energy contributions
\begin{equation}
    F\left[\boldsymbol{\rho}(\rb)\right]=F^\mathrm{ig}\left[\boldsymbol{\rho}(\rb)\right]+F^\mathrm{hs}\left[\boldsymbol{\rho}(\rb)\right]+F^\mathrm{hc}\left[\boldsymbol{\rho}(\rb)\right]+F^\mathrm{disp}\left[\boldsymbol{\rho}(\rb)\right]
\end{equation}
with repulsive hard-sphere interactions\cite{rosenfeld1997fundamental, roth2002fundamental, yu2002structures} (hs), hard-chain formation\cite{tripathi2005microstructure1,tripathi2005microstructure2} (hc), and van der Waals (dispersive) attraction of chain fluids \cite{GrossSadowski2001, sauer2017classical} (disp).
The White-Bear functional\cite{roth2002fundamental,yu2002structures} is based on Rosenfeld's Fundamental Measure Theory \cite{rosenfeld1989free} and is a commonly used Helmholtz energy functional to model hard sphere repulsion.
However, we find the functional inadequate for the description of fluids in the narrow cylindrical pores encountered in the COF frameworks.
Rosenfeld \cite{rosenfeld1997fundamental} presented a modification to Helmholtz energy functionals based on Fundamental Measure theory for fluids in strong confinement that reduces the effective dimensionality of the system.
The resulting antisymmetrized functional yields accurate results for hard spheres in narrow cylindrical pores, e.g. quasi one-dimensional systems, while retaining the full three-dimensional properties and the bulk behaviour of the original White Bear functional.
Additional details on the hard-sphere functional used in this work is provided in the Supporting Information.

The required pure component parameters for the utilized Helmholtz energy contributions are the number of segments per molecule $m_i$, the segment size parameter $\sigma_i$ and the dispersive energy parameter $\varepsilon_i$. We here use an approach that does not capture the connectivity of the different segments of a chain. Rather, the local density of segments $\rho_i(\rb)$ are considered as averages over all segments $\alpha_i$ of the chain, as
\begin{equation}
    \rho_i(\rb)=\frac{1}{m_i}\sum_{\alpha_i}^{m_i}\rho_{\alpha_i}(\rb)
    \label{eq:average_rho}
\end{equation}
leading to $\rho_i(\rb)=\rho_{\alpha_i}(\rb)$ for homosegmented chains.

The external potential represents the van der Waals interactions exerted by the COF atoms onto a fluid (segment). The external potential is calculated by considering the interactions of a PC-SAFT molecule with all individual solid atoms of the framework, leading to 
\begin{equation}
    V_i^\mathrm{ext}\left(\rb\right)=m_i \sum_{\alpha=1}^M 4\varepsilon_{\alpha i}\left(\left(\frac{\sigma_{\alpha i}}{\left\lvert \rb_\alpha-\rb \right\rvert}\right)^{12}-\left(\frac{\sigma_{\alpha i}}{\left\lvert \rb_\alpha-\rb \right\rvert}\right)^{6}\right)
\end{equation}
where $M$ is the number of solid atom interaction sites of the considered framework and $\rb_\alpha$ is the position of the atom interaction site $\alpha$ generated from the Crystallographic Information File (CIF) of the COF framework. The interaction parameters $\varepsilon_{\alpha i}$ and $\sigma_{\alpha i}$ are calculated using Lorentz-Berthelot combining rules\cite{Lorentz1881, Berthelot1898} with
\begin{align*}
    \sigma_{\alpha i}&=(\sigma_\alpha+\sigma_i)/2\\
    \varepsilon_{\alpha i}&=\sqrt{\varepsilon_\alpha \varepsilon_i}
\end{align*}
where $\sigma_\alpha$ and $\varepsilon_\alpha$ are the Lennard-Jones interaction parameters of atom interaction site $\alpha$ taken from the DREIDING force field\cite{mayo1990dreiding} representing the COF structure.

In this work, the vector containing the number of adsorbed molecules $\mathbf{N}^\mathrm{ads}=\left\lbrace N_i^\mathrm{ads},i=1,\dots,\nu\right\rbrace$ of a $\nu$ component mixture is calculated with
\begin{equation}
    \mathbf{N}^\mathrm{ads}=\int \boldsymbol{\rho}\left(\rb\right)\mathrm{d}\rb
    \label{eq:N_ads}
\end{equation}
using the vector of density profiles $\boldsymbol{\rho}\left(\rb\right)$ of all components in the system. Similar to experiments, the fluid in the COF framework is in equilibrium with a bulk phase reservoir. The number of adsorbed molecules $\mathbf{N}^\mathrm{ads}$ in the COF framework can then be calculated from the bulk conditions: for defined temperature $T$, pressure $p$ and molar fractions $\mathbf{x}=\left\lbrace x_i,i=1,\dots,\nu\right\rbrace$ of the bulk reservoir, we first calculate the chemical potentials $\boldsymbol{\mu}(p,T,\mathbf{x})$ from the PC-SAFT equation of state, we then use eq.~\eqref{eq:euler_lagrange} for determining the equilibrium densities $\boldsymbol{\rho(\rb)}$ and subsequently obtain the adsorbed amount using eq.~\eqref{eq:N_ads}.

\subsection{Grand Canonical Monte Carlo Simulation}
%
All GCMC calculations were performed using the molecular simulation software RASPA \cite{Dubbeldam2016}. 
Intramolecular fluid and intermolecular fluid-fluid interactions were described with the TraPPE force field.\cite{Siepmann1998, Siepmann2004} 
The CH$_\mathrm{x}$ groups in methane, ethane and $n$-butane were considered as single, chargeless interaction centers (united atoms) 
with effective Lennard-Jones potentials. 
Parameters for unlike interaction sites were determined 
using Lorentz-Berthelot combining rules. 
TraPPE approximates the quadrupolar nature of nitrogen by placing negative partial 
atomic charges at the position of the nitrogen atoms and a neutralizing positive partial charge at the center of mass. 
The COF framework was considered to be rigid such that only Lennard-Jones parameters and partial atomic charges needed to be assigned to the different 
atomic species. The Lennard-Jones parameters were taken from the DREIDING force field\cite{mayo1990dreiding}.
Partial atomic charges of the COF structures were calculated using the extended charge equilibration (EQeq)\cite{Willmer2012} method implemented in RASPA. EQeq expands charge equilibration (Qeq)\cite{Rappe1991} including measured ionization energies. The method was tested for screening MOFs\cite{Willmer2011} and is computationally fast. For the purpose of the present work, where partial charges play only a minor role, this approach is sufficient. For other purposes an evaluation of different variants of the algorithm\cite{Ongari_JCTC2019} may be required or training the algorithm for COFs\cite{Woo2013,Deeg_ACSAMI_2020}. Also test calculations using sophisticated methods such as REPEAT\cite{Campagna2009} or DDEC \cite{Manz2010} which are based on electronic structure calculations on the cDFT level are recommended to validate results from EQeq calculations.
All force field parameters applied in the present work are reported in the Supporting Information. 
To be comparable to the classical DFT calculations described in the previous section, the cut-off radius used for the Lennard-Jones corrections 
was 14.816 \AA\:, 
which is equal to four times the $\sigma_\mathrm{CH_4}$-parameter of the 
PC-SAFT EoS\cite{GrossSadowski2001}. Although the 
density beyond the cut-off radius is not uniform, we applied analytic corrections to the long-range Lennard-Jones tail, in order to reduce the sensitivity of the results with respect to
the cut-off radius\cite{Jablonka_2019}.
The real part of the electrostatic interactions was evaluated up to a cut-off radius of  12.0 \AA.
Long-range electrostatic interactions were calculated by Ewald summation\cite{Ewald1921, Dubbeldam2013} 
with a relative precision of $10^{-6}$. 
To carry out simulations at constant chemical potential, the PC-SAFT EoS was used to pre-compute a fugacity 
coefficient at the given temperature and pressure that was then passed to the MC code. The number of MC cycles was $25\times10^4$,
both, for equilibration and for the production phase. One cycle consists of $\max(20, N_t)$ MC moves (with $N_t$ as the sum of adsorbate molecules in the system),
i.e. translation, insertion or 
deletion and, in case of molecules represented by more than one site, rotation moves. In simulations of binary mixtures 
identity swap moves were carried out additionally. All moves were performed with equal probability.  

\subsection{COF Structures} 
\begin{figure}[bht]
    \centering
    \includegraphics[width=0.7\textwidth]{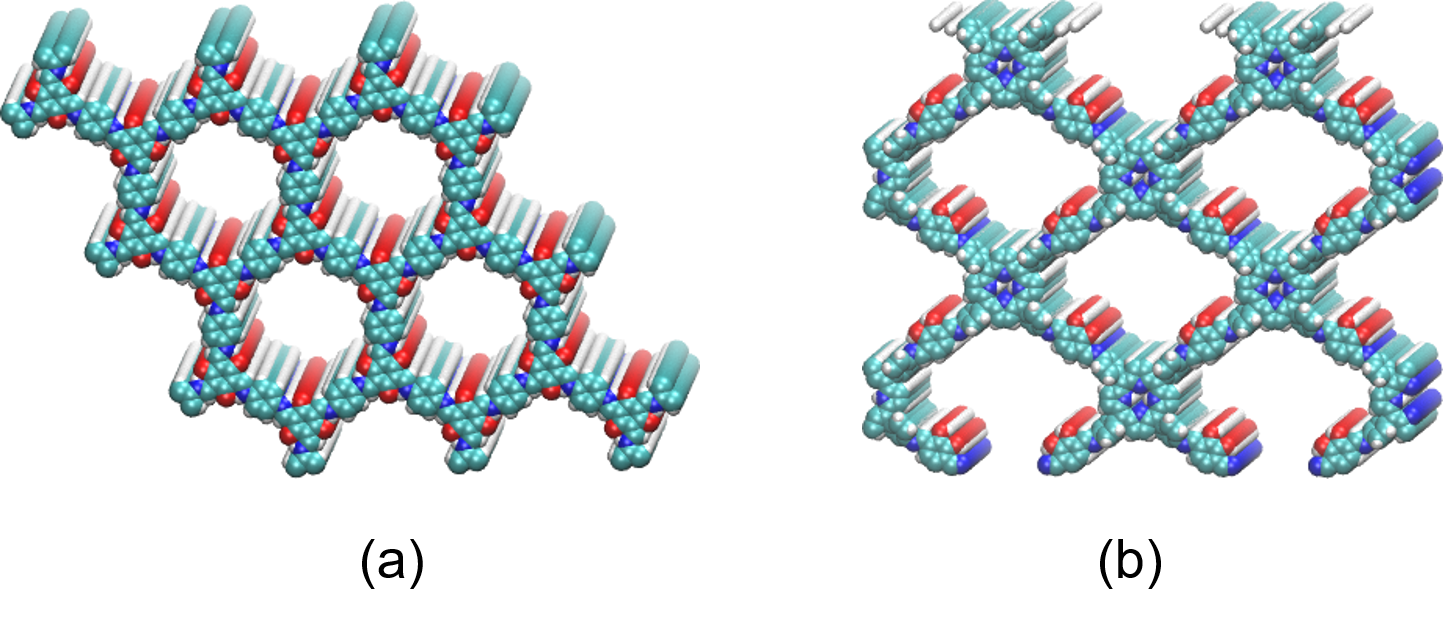}
   \caption{Structural represenation of the covalent organic frameworks studied in the present work. (a) TpPa-1; (b) 2,3-DhaTph. Carbon, nitrogen, oxygen and hydrogen are represented as cyan, blue, red and white spheres, respectively.}
    \label{fgr:frameworks}
\end{figure}

The two COFs considered in the present work are the ketoenamine-linked COF TpPa-1\cite{Kandambeth2012} and the imine-linked COF 2,3-DhaTph\cite{Kandambeth_2013,C4CC07104B} having pore sizes of approximately 1.8 and 2.0~nm, respectively, see Figure~\ref{fgr:frameworks}.

As the purpose of this study is the comparison between two computational approaches somewhat idealized structures were used. For the hexagonal COF TpPa-1 we assumed a perfectly eclipsed arrangement, with coordinates taken from Cambridge Structural Database\cite{Groom2016} under deposition number 945096\cite{Kandambeth2012}. A detailed investigation of the effects of interlayer slipping on adsorption was for example reported by Sharma et al.\cite{C7CE01647F}

For the COF 2,3-DhaTph initial coordinates in a perfectly eclipsed arrangement were taken from the CoRe COF database\cite{TONG2017456,TONG2018}. However, the layer-layer distance in that structure of 6.7~\AA~is much larger than the experimentally reported value of 4.0~\AA~because the benzene rings were rotated by 90$^{\circ}$. 
Moreover, the lattice was not tetragonal as in the experimentally derived X-ray structure,\cite{Kandambeth_2013,C4CC07104B} but rather orthorhombic.
To avoid artificial adsorption of adsorbates between the layers the benzene rings were rotated by approximately 30$^{\circ}$ resembling the value in the experimental structure, which allowed to bring the layers closer together to 4.0~\AA~ in our computational model without inducing steric clashes.
In the GCMC simulations 9 layers were used for TpPa-1 and 8 layers for 2,3-DhaTph, resulting in simulation box sizes of 3.06 and 3.2~nm in $z$-direction, respectively. For the rectangular box of 2,3-DhaTph, the other dimensions are 4.0028 and 3.259~nm and for the hexagonal box of TpPa-1 the lengths are 4.5112~nm in each direction.
The CIF-files of the two structures used in the present work are provided as Supporting Information.

\subsection{Ideal Adsorbed Solution Theory}



Adsorption isotherms of mixtures can be estimated from the pure component isotherms using the ideal adsorbed 
solution theory (IAST).\cite{Myers1965}
In the present work the IAST equations were solved with the pyIAST package.\cite{pyIAST2016}
To account for non-ideal behavior of the gas phase at elevated pressure fugacities instead of pressures were employed in the IAST equations.\cite{Keil2010,Krishna2018}

\section{Results and Discussion}


Before comparing adsorption isotherms predicted by cDFT and GCMC we first investigate vapor-liquid equilibria 
to assess whether the two approaches show deviations that may impact their comparability. 
Note that the segment size parameter $\sigma_{ii}$ and the dispersive energy parameter $\varepsilon_{ii}$ used in PC-SAFT 
are different from the force-field parameters used in the MC simulations, even for methane. Both were independently adjusted to experimental data of pure compounds. 
Results from both approaches are comparable, however, because pure component parameters were adjusted to experimental data for phase equilibria. 

\begin{figure}[ht]
    \centering
    \includegraphics[width=1.0\textwidth]{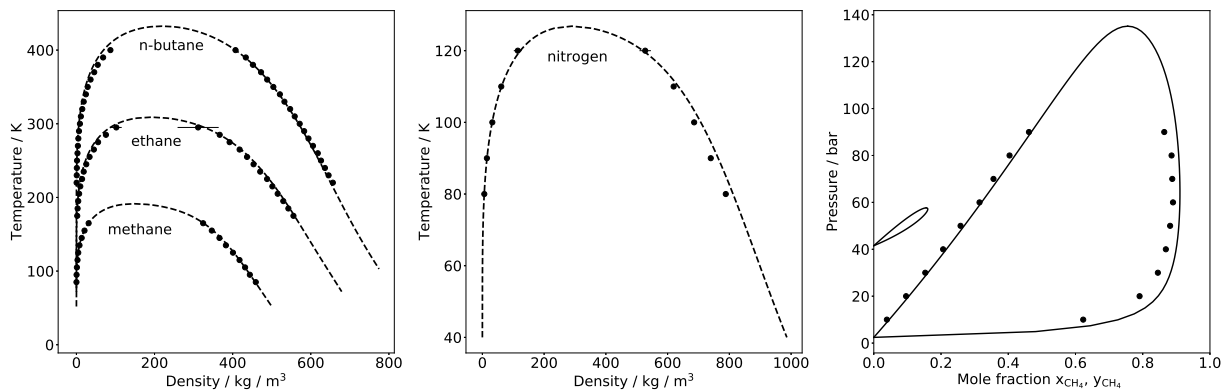}
   \caption{Vapor-liquid coexistance curves for (a) methane, ethane, n-butane, (b) nitrogen and (c) binary mixtures of methane/ethane and methane/$n$-butane. For the binary mixtures the equilibrium pressure is plotted over the methane mole fraction at 298 K. The symbols represent Gibbs-Ensemble Monte Carlo simulations, the lines PC-SAFT calculations.}
    \label{fgr:phase_diagrams}
\end{figure}

Figure~\ref{fgr:phase_diagrams}
shows that vapor-liquid coexistance curves in the temperature-density projection for nitrogen, methane, ethane and $n$-butane 
obtained from Gibbs-Ensemble\cite{Panagiotopoulos_1987,Panagiotopoulos_1988} Monte Carlo simulations do not exhibit significant deviations between the two methods. 
For the vapor liquid equilibrium of the mixtures some deviations occur for the methane/$n$-butane system. These deviations in the vapor phase can be attributed to rather significant deviations in vapor pressures observed for the TraPPE force field\cite{hemmen2015transferable}.
For the mixture of methane/ethane sampling of a stable two-phase region was difficult to establish with Gibbs Ensemble Monte Carlo, because the vapor-liquid phase envelop is rather small and the system is close to the mixtures' critical point for all relevant compositions. 
However, simulations at 199.93~K, reported by Chakraborti and Adhikari\cite{Chakraborti_2017} showed a good agreement with experiment for the saturated liquid phase but significant deviations in the coexisting vapor phase, similar to the methane/$n$-butane case studied here. As shown below, these differences do not have a significant impact on the adsorption equilibria in the considered pressure range. Therefore, an attempt to
use improved variants of the TraPPE force field\cite{Shah_2017} was not pursued.


\subsection{Pure Component Adsorption}
\begin{figure}[ht]
    \centering
    \includegraphics[width=0.5\textwidth]{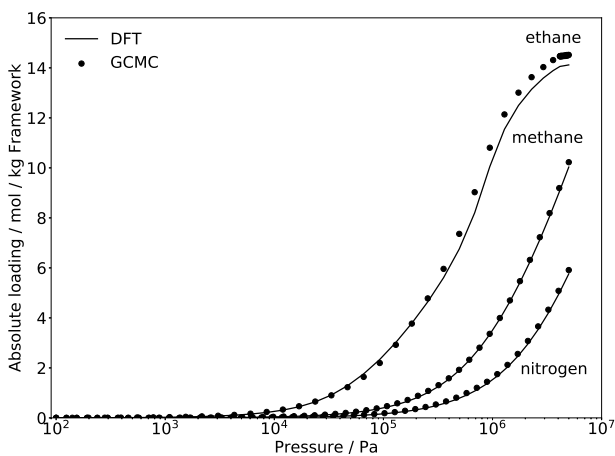}
   \caption{Pure component adsorption isotherms of nitrogen, methane and ethane in COF 2,3-DhaTph at 298 K obtained from GCMC simulations and classical DFT calculations.}
    \label{fgr:pure_all_dha}
\end{figure}

The pure component adsorption isotherms are presented by plotting the average absolute amount adsorbed $N^\mathrm{ads}$ per mass of COF-framework as function of the pressure in the external reservoir. 
The statistical uncertainties in the GCMC results are in almost all cases smaller than the symbol size. 
Figure~\ref{fgr:pure_all_dha} shows adsorption isotherms of nitrogen, methane and ethane in COF 2,3-DhaTph at 298 K with varying pressure
up to 50 bar. For methane excellent agreement between cDFT and GCMC is obtained over the entire pressure range. 
For ethane the agreement between the two approaches is very good up to pressures of 2 bar. At higher pressures cDFT slightly underestimates 
the amount adsorbed.
For nitrogen the agreement between GCMC and cDFT is remarkable given that the force field contains three collinear partial atomic 
charges to model the quadrupolar nature of the molecule while the PC-SAFT model entering the cDFT calculations 
describes nitrogen as non-polar, so that the van der Waals parameters effectively capture the (mild) quadrupole moment of nitrogen.
Of course, the influence of the partial atomic charges on the adsorption behavior strongly depends on the magnitude of the 
charges of the adsorbent-framework, as reported for siliceous zeolites.\cite{HACKETT2018231} 
For here considered COF 2,3-DhaTph nitrogen adsorption 
isotherms with and without framework partial charges show only minor differences (see Figure S7 in the Supporting Information).

\begin{figure}[ht]
    \centering
    \includegraphics[width=0.5\textwidth]{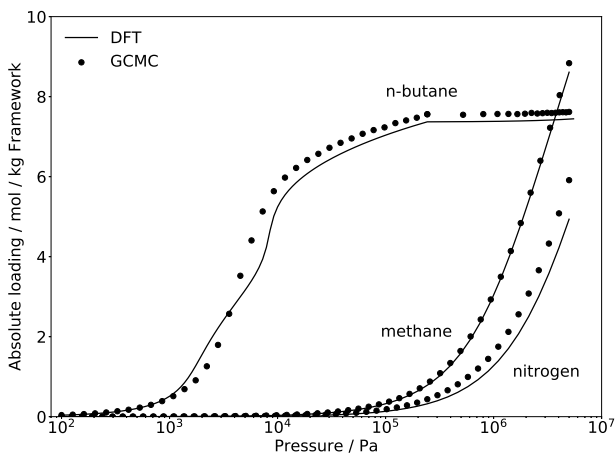}
    \caption{Pure component adsorption isotherms of nitrogen, methane and $n$-butane in COF TpPa-1 at 298 K obtained from GCMC simulations and classical DFT calculations.}
    \label{fgr:pure_all_tppa}
\end{figure}

Figure~\ref{fgr:pure_all_tppa} shows adsorption isotherms of nitrogen, methane and n-butane in COF TpPa-1 at 298~K up 
to a pressure of 50 bar. As before, excellent agreement between GCMC and cDFT is obtained for methane. For n-butane 
some deviations occur in the pressure range between $10^3$ and $10^4$ Pa in which the shape of the cDFT 
isotherm is less smooth compared to its GCMC counterpart, possibly due to the averaged segment-density according to eq.~\eqref{eq:average_rho}. Based on these deviations, it is interesting to investigate a cDFT functional, where connectivity of segments is accounted for and the densities of individual segments are calculated. The formalism was proposed by Jain and Chapman\cite{jain2007modified} and has also been applied with the PC-SAFT DFT model in previous work of our group\cite{mairhofer2018classical}.
For nitrogen the cDFT isotherm is slightly lower than the GCMC one. Again, we tested the influence of the partial charges on the framework atoms with respect to nitrogen adsorption and found that the GCMC isotherm is in excellent agreement with the cDFT isotherm when evaluated in a framework exempt from partial charges (see Supporting Information).
The TpPa-1 framework is somewhat more polar than 2,3-DhaTph, 
if we use the sum of squared partial charges $q_i$ as a measure for how polar a framework is. We thus regard the sum of $N_iq_i^2/V$, where $N_i$ is the number atoms of species $i$ in the simulation cell and $V$ its volume. For all of the four atomic species (C, H, N, O) these values are higher for TpPa-1 compared to 2,3-DhaTph.
Therefore, 
for frameworks with low to moderate charge densities,
adsorption of quadrupolar fluids may be approximated by dispersion interactions alone. 
In summary, and in view of the fact that no parameter is adjusted for relating the two modelling approaches, we consider the overall agreement observed in Fig.~\ref{fgr:pure_all_dha} and \ref{fgr:pure_all_tppa} as good.

\subsection{Binary Mixture Adsorption}
\begin{figure}[ht!]
    \centering
    \includegraphics[width=0.5\textwidth]{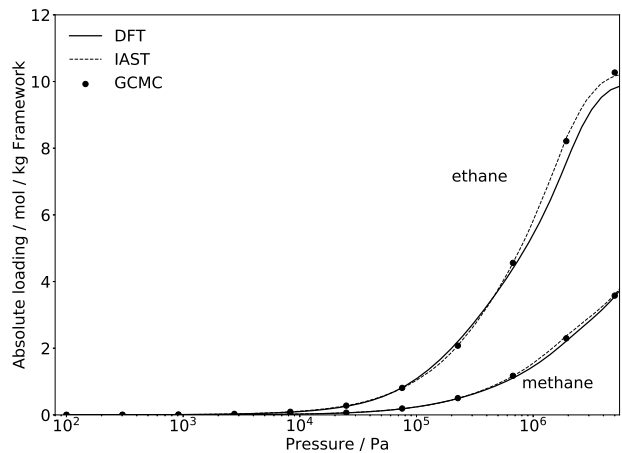}
    \caption{Adsorption isotherms for the methane/ethane mixture in COF 2,3-DhaTph at 298 K and $x^{\mathrm{bulk}}_{\mathrm{CH_4}} = 0.6$. 
    The IAST results are based on fits to the pure component GCMC isotherms.}
    \label{fgr:dha_mix_06}
\end{figure}

The binary mixture isotherms are presented by plotting the average absolute amount adsorbed of each species 
as function of the total pressure in the external reservoir. 
The mixture adsorption of methane and ethane was studied in the COF 2,3-DhaTph at 
methane mole fractions of the reservoir mixture of $x_{\mathrm{CH_4}}^{\mathrm{bulk}} = \{0.1, 0.4, 0.6, 0.8\}$. 
Figure~\ref{fgr:dha_mix_06} shows the case $x_{\mathrm{CH_4}}^{\mathrm{bulk}} = 0.6$. 
All other cases are presented in the Supporting Information. 
The adsorption isotherms of methane and ethane in the mixture predicted by cDFT are in very good agreement with the GCMC results. Only at higher pressure GCMC predicts a slightly larger amount adsorbed of ethane, as can be expected from the results obtained from the pure component isotherms discussed above. IAST is in very good agreement with the GCMC results indicating that the adsorbed phase is approximated well by an ideal solution. We note, however, that IAST takes the results from GCMC simulations of pure substances as input.

\begin{figure}[ht!]
    \centering
    \includegraphics[width=0.5\textwidth]{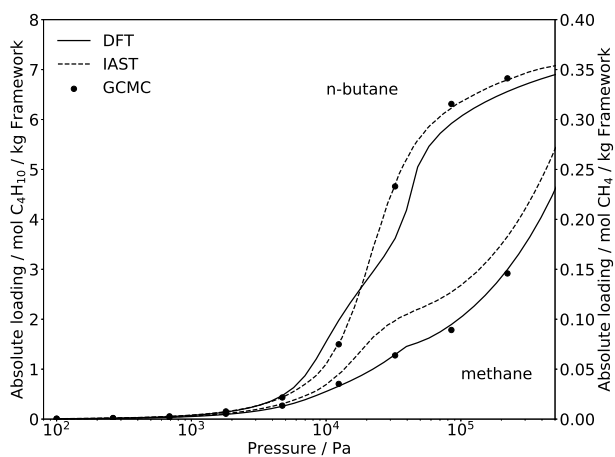}
    \caption{Adsorption isotherms for the methane/$n$-butane mixture in COF TpPa-1 at 298 K and $x^{\mathrm{bulk}}_{\mathrm{CH_4}} = 0.8$. 
    The IAST results are based on fits to the pure component GCMC isotherms.
    Methane adsorption is displayed on the secondary y-axis with higher resolution.}
    \label{fgr:tppa_mix_08}
\end{figure}

The mixture adsorption of methane and $n$-butane was studied in the COF TpPa-1 at methane mole fraction
of $x_{\mathrm{CH_4}}^{\mathrm{bulk}} = \{0.2, 0.4, 0.6, 0.8\}$. Figure~\ref{fgr:tppa_mix_08} shows the case 
$x_{\mathrm{CH_4}}^{\mathrm{bulk}} = 0.8$, all other cases are presented in the Supporting Information.
%
Due to the much stronger adsorption of $n$-butane relative to methane we introduced a second $y$-axis in Fig.~\ref{fgr:tppa_mix_08}, to better present the
amount of adsorbed methane. For methane, cDFT and GCMC are in very good agreement as can be expected from the 
comparison of the pure methane isotherms discussed above. The $n$-butane isotherms as predicted from cDFT underestimate the adsorbed amount as compared to the GCMC results, similar to the behavior for pure $n$-butane. For methane IAST predicts a slightly higher amount adsorbed above pressures of $10^4$~Pa.



\subsection{Conclusion}
A classical DFT approach relying on a Helmholtz energy functional based on the PC-SAFT equation 
of state was used to predict adsorption equilibria of pure components and binary mixtures in 
covalent organic frameworks. The results were compared to adsorption isotherms from GCMC simulations (using the TraPPE force field for the fluids). 
While the latter approach is rooted in a fully atomistic description (within a united atom 
approximation for alkanes), cDFT employs a coarser description of the fluid by means of an analytical 
equation of state. The basis of the comparison is thus the ability of both approaches to describe 
the properties of the bulk fluid phases. Regarding the adsorption, both approaches employ the same
Lennard-Jones parameters for the atoms of the COF-framework (DREIDING force field). Solid-fluid
interactions are defined using Berthelot-Lorentz combining rules.
In the case of cDFT using the PC-SAFT functional, however, van der Waals segment size and energy
parameters enter the combining rules, whereas in the case of GCMC the Lennard-Jones parameters of each interaction 
site of the TraPPE model enter the combining rules. The remarkable agreement of the two approaches, even for $n$-butane, shows that cDFT is a powerful alternative to GCMC for studying adsorption equilibria in porous materials.
The advantage of this approach is its analytical nature allowing the calculation of derivatives and, therefore, optimization tasks with respect to force field parameters or structural properties of the porous materials. 
The second aspect is in particular relevance for COFs because the stacking motifs have a substantial impact on the 
materials properties, including the adsorption behavior. The current limitation of the cDFT approach is a 
lower coverage of the chemical space compared to molecular simulations, in particular with regard to polar molecules.

\section{Author Notes}
Christopher Kessler and Johannes Eller contributed equally to this work.

\begin{acknowledgement}
This work was funded by the Deutsche Forschungsgemeinschaft (DFG, German Research Foundation) - Project-ID 358283783 - SFB 1333 and Project-ID 327154368 – SFB 1313. Monte Carlo simulations were performed on the computational
resource BinAC at High Performance and Cloud Computing Group at the Zentrum f\"ur Datenverarbeitung of the University
of T\"ubingen, funded by the state of Baden-W\"urttemberg through bwHPC and the German Research Foundation (DFG) through grant no INST 37/935-1 FUGG.
\end{acknowledgement}

\begin{suppinfo}


Further details on the cDFT calculations and Gibbs-ensemble MC simulations, Force-field parameters, additional simulation results for mixtures, CIF files of the COF-structures, input files for RASPA and a Jupyter notebook containing the data analysis are found the supporting material.

\end{suppinfo}

\appendix

\bibliography{GCMC_DFT_COF}

\end{document}